\documentclass[fleqn,10pt]{wlscirep}
\usepackage[utf8]{inputenc}
\usepackage[T1]{fontenc}
\title{Structural transition in social networks: The role of homophily}

\author[1,*]{Yohsuke Murase}
\author[2,3,4]{Hang-Hyun Jo}
\author[5,6,7]{J\'anos T\"or\"ok}
\author[6,5,4,+]{J\'anos Kert\'esz}
\author[4,8]{Kimmo Kaski}

\affil[1]{RIKEN Center for Computational Science, Kobe, Hyogo 650-0047, Japan}
\affil[2]{Asia Pacific Center for Theoretical Physics, Pohang 37673, Republic of Korea}
\affil[3]{Department of Physics, Pohang University of Science and Technology, Pohang 37673, Republic of Korea}
\affil[4]{Department of Computer Science, Aalto University, Espoo FI-00076, Finland}
\affil[5]{Department of Theoretical Physics, Budapest University of Technology and Economics, Budapest H-1111, Hungary}
\affil[6]{Department of Network and Data Science, Central European University, Budapest H-1051, Hungary}
\affil[7]{MTA-BME Morphodynamics Research Group, Budapest University of Technology and Economics, Budapest H-1111, Hungary}
\affil[8]{The Alan Turing Institute, British Library, London NW1 2DB, UK}
\affil[*]{yohsuke.murase@gmail.com}
\affil[+]{KerteszJ@ceu.edu}
\begin{abstract}
We introduce a model for the formation of social networks, which takes into account the homophily or the tendency of individuals to associate and bond with similar others, and the mechanisms of global and local attachment as well as tie reinforcement due to social interactions between people. We generalize the weighted social network model such that the nodes or individuals have $F$ features and each feature can have $q$ different values. Here the tendency for the tie formation between two individuals due to the overlap in their features represents homophily. We find a phase transition as a function of $F$ or $q$, resulting in a phase diagram. For fixed $q$ and as a function of $F$ the system shows two phases separated at $F_c$. For $F{<}F_c$ large, homogeneous, and well separated communities can be identified within which the features match almost perfectly (segregated phase). When $F$ becomes larger than $F_c$, the nodes start to belong to several communities and within a community the features match only partially (overlapping phase). Several quantities reflect this transition, including the average degree, clustering coefficient, feature overlap, and the number of communities per node. We also make an attempt to interpret these results in terms of observations on social behavior of humans.
\end{abstract}
\begin{document}

\flushbottom
\maketitle
\thispagestyle{empty}

%\noindent Please note: Abbreviations should be introduced at the first mention in the main text – no abbreviations lists. Suggested structure of main text (not enforced) is provided below.

\section*{Introduction}\label{sec:intro}
In human societies homophily, the tendency of similar individuals getting associated and bonded with each other, is known to be a prime tie formation factor between a pair of individuals~\cite{mcpherson2001birds}. This association and bonding can be related to one or more features including gender, race, age, education level, economic and social status, and many more. Consequently homophily has a large impact on a number of fundamentally important social phenomena like segregation, inequality, perception biases, and the transmission of information between groups of individuals~\cite{schelling1969models, Karimi2018Homophily, lee2017homophily, Halberstam2016information}. The mechanism of focal closure, i.e., the process of forming links to others with shared characteristics but without common acquaintances can be considered as a typical manifestation of homophily~\cite{kossinets2006empirical, kumpula2007emergence}.

The system that emerges due to inter-personal tie formation, constitutes a social network, which is one of the paradigmatic examples of complex networks~\cite{Newman2010networks, Barabasi2016networksci}. Noticeable features of social networks are the significantly high occurrence of triangles (clustering) and the structure of communities~\cite{hric2016community}, which are the densely ``wired'' parts of the network as compared to their connectivity to the rest of the network. Communities themselves have a complex structure, characterized by overlaps, hierarchy, and multiplexity~\cite{palla2005uncovering, girvan2002community, Lancichinetti2009detecting, ahn2010link, rosvall2008maps, rosvall2011multilevel, murase2015modeling}. Moreover, the topological features are correlated with the intensity of the ties represented by the weights of the links. The ``strength of weak ties'' hypothesis proposed by Granovetter~\cite{granovetter1973strength}, stating that a stronger tie between two persons leads to larger overlap between their neighbors, has been validated on empirical data~\cite{onnela2007analysis, onnela2007structure}. It suggests that the communities consist of strong links while inter-community links are weak. Studies have shown that microscopically a key inducing factor of such community structure is the mechanisms of cyclic closure~\cite{kossinets2006empirical, kumpula2007emergence}, the tendency of nodes in social networks to make links with a topologically close nodes, i.e., people often form social ties with a person sharing common friends.

The two main mechanisms for social tie formation, the focal and cyclic closure, amplify the similarity of nodes in a community~\cite{kossinets2009origins, mcpherson2001birds}. As said above the focal closure is considered being based on homophily, while the triadic closure means the tie formation between two acquaintances of an ego, when these meet at an occasion usually related to similarity in values or features. Thus even a relatively weak preference for homophilic relationships would tend to be amplified over time, via a cumulative advantage. As Kossinets and Watts discussed~\cite{kossinets2009origins}, this casts the question: To what extent can observed patterns of homophily be attributed to individual preferences and structural constraints? They also discussed that ``A thorough answer to this question would require the use of simulation models, in which choice homophily as well as focal and cyclic closure biases could be systematically varied.'' The purpose of this paper is to contribute to answering this question by studying the macroscopic consequences of the joint effect of focal and cyclic closure mechanisms.

To this end, we develop a model based on the weighted social network (WSN) model~\cite{kumpula2007emergence}. In the WSN model two main mechanisms for the tie-formation process were used, namely local attachment (LA) and global attachment (GA), corresponding to triadic and focal closure, respectively, together with link strength reinforcement. With this minimal model the complex Granovetterian weight-topology relation of social networks could be successfully reproduced~\cite{kumpula2007emergence}. More recently, this model was generalized to temporal and multiplex networks~\cite{jo2011emergence, murase2014multilayer, murase2015modeling}.

In this paper the WSN model is generalized to enable homophilic relationships. We introduce features of the individuals (represented by the nodes of the network) in the spirit of the classical Axelrod model~\cite{axelrod1997dissemination} by modeling the different traits of a node with components of an $F$-dimensional vector. The Axelrod model was intensively investigated by San Miguel and coworkers~\cite{centola2007homophily, vazquez2007time, min2017fragmentation, Battiston2017layered} focusing mainly on the problem of cultural drift and fragmentation by applying dynamic rules to describe the changes of the features of the nodes upon interactions. In most cases fixed network geometries were applied, except in Ref.~\cite{min2017fragmentation}, where the effect of network adaptation was studied. Here we take a different approach by considering the features fixed and model the tie formation based on homophilic interactions as suggested by social network theory. We show that homophily has a major impact on the emerging network structure. Depending on the number $F$ of features there is a transition from a few large and homogeneous communities to many smaller and heterogeneous ones. The possible number $q$ of different values a feature can take is also a relevant parameter and we show that a scaling relationship between the average degree of nodes, $F$, and $q$ is approximately valid.

The paper is organized as follows: After we define the model in the next section, we demonstrate that the model exhibits two regimes depending on the parameter values. We also show that the similarity between nodes is very much amplified in one of the phases, and show that a global segregation is observed. Finally, we present a summary and discussion in the last section.

\section*{Model}\label{sec:model}

First we briefly review the original weighted social network (WSN) model proposed in Refs.~\cite{kumpula2007emergence, murase2015modeling}, which is a dynamic, undirected, weighted network model leading to a stationary state. We start with $N$ unconnected nodes and links between them are created and updated with the following three mechanisms:
\renewcommand{\labelenumi}{\roman{enumi}}
\begin{enumerate}
\item \textit{global attachment} (GA): A node is selected at random, its degree is $k$. With probability $\delta_{0,k}+(1-\delta_{0,k})p_r$ it is connected with a new link of weight $w_{0}$ to a randomly chosen node. In the unlikely event that the two nodes are already connected, a new target node is chosen.
\item \textit{local attachment} (LA): A randomly chosen node $i$ (with $k_i\neq0$) chooses one of its neighbors $j$ with the probability proportional to $w_{ij}$ which stands for the weight of the link between nodes $i$ and $j$.
Then node $j$ chooses one of its neighbors but $i$, say $k$, randomly with probability proportional to $w_{jk}$ and if nodes $i$ and $k$ are not connected, they are connected with probability $p_{\Delta}$ with a link of weight $w_{0}$.
In addition, all the involved links increase the weights by $w_r$, whether a new link is created or not.
\item \textit{link deletion} (LD): Each link is removed from the system with probability $p_{d}$ at each time step. (Time is measured in sweeps.)
\end{enumerate}
Sequential update is applied, first GA, then LA to the nodes and then LD to the links. Over the wide range of the parameters this model produces Granovetterian structures with strongly wired communities connected by weak ties~\cite{kumpula2007emergence, murase2015modeling}.

In order to incorporate the effect of homophily into the model we extend the WSN model in the following way. Similarly to the Axelrod model and its variants~\cite{axelrod1997dissemination, centola2007homophily, vazquez2007time, vazquez2007non, gandica2011cluster, tilles2015diffusion, min2017fragmentation}, we assign a vector of $F$ components to each node $i$, corresponding to $F$ different features of the node (related e.g. to gender, ethnicity, language, religion, etc.). Each feature can take one of $q$ possible values. Then the feature vector of node $i$ is a set of $F$ integers: $(\sigma_i^1,\sigma_i^2,\dots,\sigma_i^F)$ with $\sigma_i^f\in\{1,\cdots,q\}$ for each $f\in\{1,\cdots,F\}$. The size of the vector, $F$, represents the social complexity of the population, i.e., the larger $F$ implies the greater number of cultural characteristics that are attributable to each individual. The number of traits per feature, $q$, represents the heterogeneity of the population: The larger $q$ means the greater number of cultural options in the society. Although $F$ may depend on individuals and $q$ may be different for each feature in reality, for simplicity we assume all nodes to have the same $F$ and keep $q$ constant. These will be parameters of our model.

The feature values or traits of the nodes are initially chosen to be uniformly random and then kept unchanged. This set-up is different from those in studies of the Axelrod model~\cite{axelrod1997dissemination, centola2007homophily, vazquez2007time}, where traits change as a consequence of interaction with similar neighbors. People in reality have both changeable and rigid traits, and the degree of the changeability may also depend on the individual and cultural properties. In this paper we study the case with fixed traits as it is the simplest setting to provide a baseline for better understanding of the impact of homophily on the network structure. We will show that segregation may occur even without a trait changing mechanism, just as a consequence of tie formation influenced by homophily.

\begin{figure}
  \begin{center}
  \includegraphics[width=.9\textwidth]{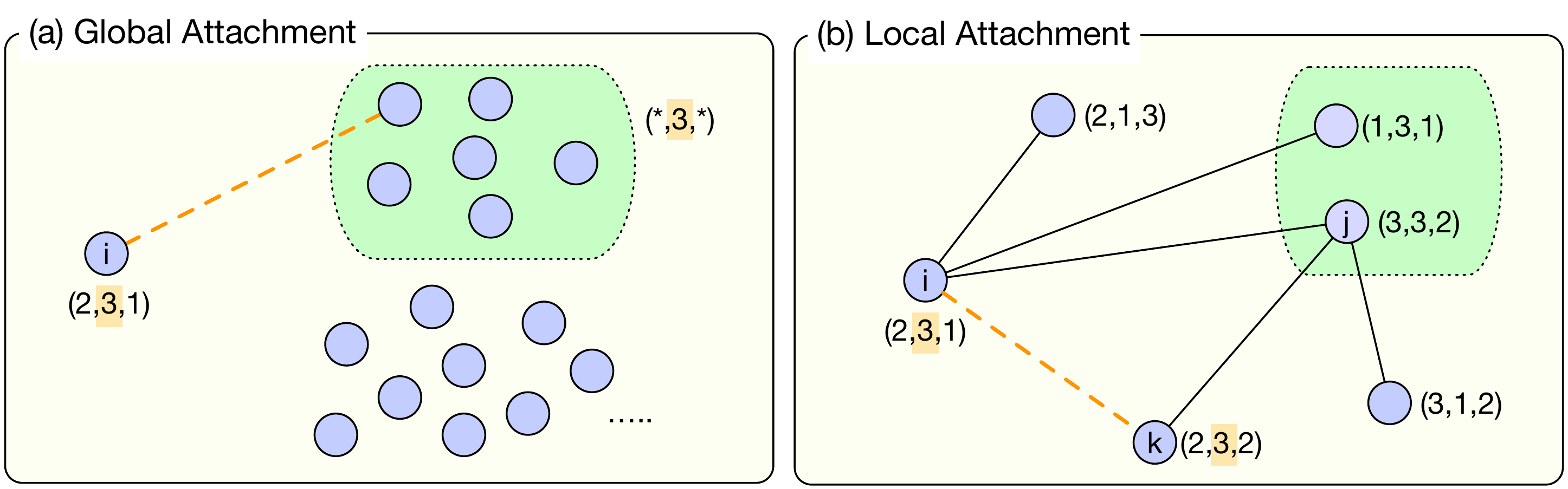}
  \caption{
  A schematic illustration of our homophilic model.
  (a) At a global attachment (GA) step, the focal node $i$ randomly chooses one of its features, $f$. Suppose that the second feature is selected, i.e., $\sigma_i^{f}=3$.
  With probability $p_r$, $i$ is connected to a node randomly chosen from the nodes having the same value in the second feature, which are indicated by the dashed square.
  (b) At a local attachment (LA) step, the focal node $i$ chooses one of its features $f$ similarly. Suppose that the second feature is selected.
  From $i$'s neighbors sharing the second feature, indicated by the dashed square, a node $j$ is selected with probability proportional to the link weight $w_{ij}$. Then $j$ chooses another node $k$ but $i$ sharing the second feature that has a chance to get connected to the node $i$.
  }
  \label{fig:model_definition}
  \end{center}
\end{figure}

In our homophilic WSN model, the network is updated similarly to the simple WSN model but the features of nodes are taken into account, as shown in Fig.~\ref{fig:model_definition}. In each GA and LA step, a feature $f$ of the focal node $i$ is randomly chosen from $F$ features and it can make links only to the nodes sharing the same trait for the feature $f$, i.e., only to the nodes $j$ satisfying $\sigma_j^f=\sigma_i^f$.

More specifically, in a GA step, the node $i$ chooses randomly a node, say $j$, sharing the same trait for the feature $f$ in the network. If the node $i$ has any neighbor sharing the same trait for the feature $f$, then the link between $i$ and $j$ is created with probability $p_{r}$. Otherwise, the link is created with probability one. The weight of the created link is given as $w_0$. In the LA step, among $i$'s neighbors sharing the same trait for the feature $f$ with $i$, choose a node, say $j$, with probability proportional to their link weights. Then, the node $j$ chooses a random node, say $k$, among $j$'s neighbors sharing the same trait for the feature $f$ but $i$ with probability proportional to the link weights. It implies that $\sigma_i^f=\sigma_j^f=\sigma_k^f$. All the links included in this triangle increase their weights by $w_r$. If nodes $i$ and $k$ are not connected, a link between them is created with probability $p_{\Delta}$ and its weight is set to $w_0$.

Note that this model reduces to the original WSN model when $q=1$, because every node pair shares the same trait, i.e., $\sigma_i^f=1$ for all $i$ and $f$. When $F=1$, the network is decomposed into $q$ disjoint subgraphs since the nodes with different traits have no chance to be connected, while the model dynamics in each subgraph is equivalent to the original WSN model but with a smaller system size of $\sim N/q$.

Starting with a network with no links, GA, LA, and LD are applied in each time step. After a sufficiently long equilibration period, ranging from $5\times 10^4$ to $8\times 10^5$ Monte Carlo steps in our work depending on the relaxation time, statistically stationary states are reached.
In the following, $N = 50000$, $p_{\Delta} = 0.02$, $p_r = 0.001$, $p_d = 0.005$, $w_r = 1$ are used.
In this study, OACIS was used for a systematic parameter scan~\cite{murase2017open}.

\begin{figure}
  \begin{center}
  \includegraphics[width=.9\textwidth]{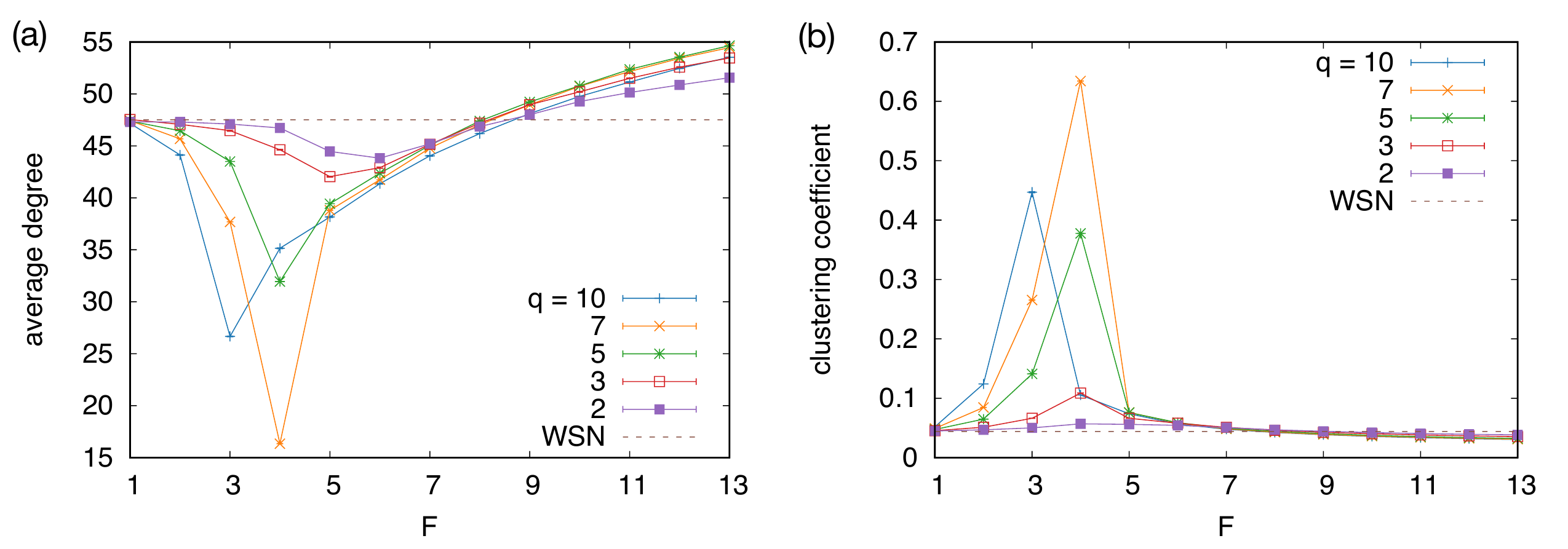}
  \caption{
  (a) Average degree and (b) clustering coefficient as a function of the number of features $F$ for different values of $q$.
  The results are averaged over five independent runs and its standard error is smaller than the symbol size.
  }
  \label{fig:k_F}
  \end{center}
\end{figure}

\begin{figure}
  \begin{center}
  \includegraphics[width=.9\textwidth]{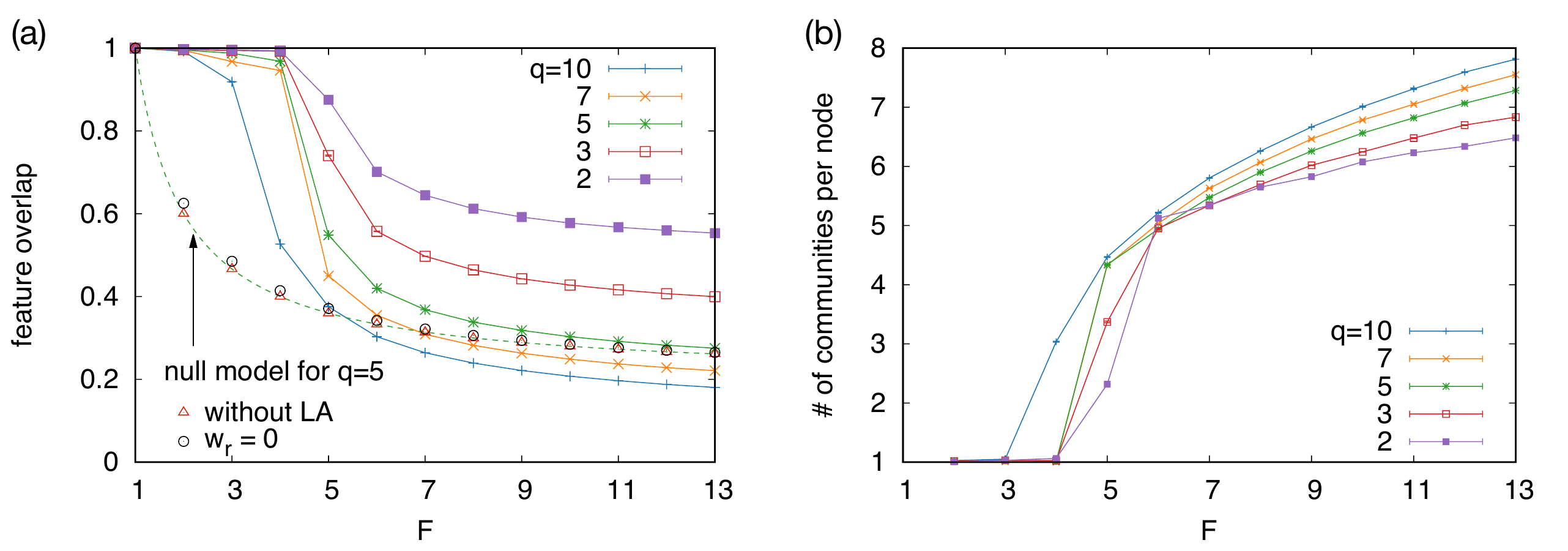}
  \caption{
  (a) Feature overlap averaged over the connected nodes are shown as a function of $F$ for several $q$.
  In addition to the simulation results, we plot a prediction (dashed line) by the null model assuming that the probability of making a link is proportional to the feature overlap between the nodes, compared to the numerical results in the case without LA steps as well as when $w_r=0$.
  (b) Average number of overlapping communities per node as a function of $F$ for various $q$.
  The hierarchical communities are detected using the multilevel Infomap method and only the communities on top of the hierarchy are taken.
  }
  \label{fig:overlap_F}
  \end{center}
\end{figure}

\begin{figure}
  \begin{center}
  \includegraphics[width=.95\textwidth]{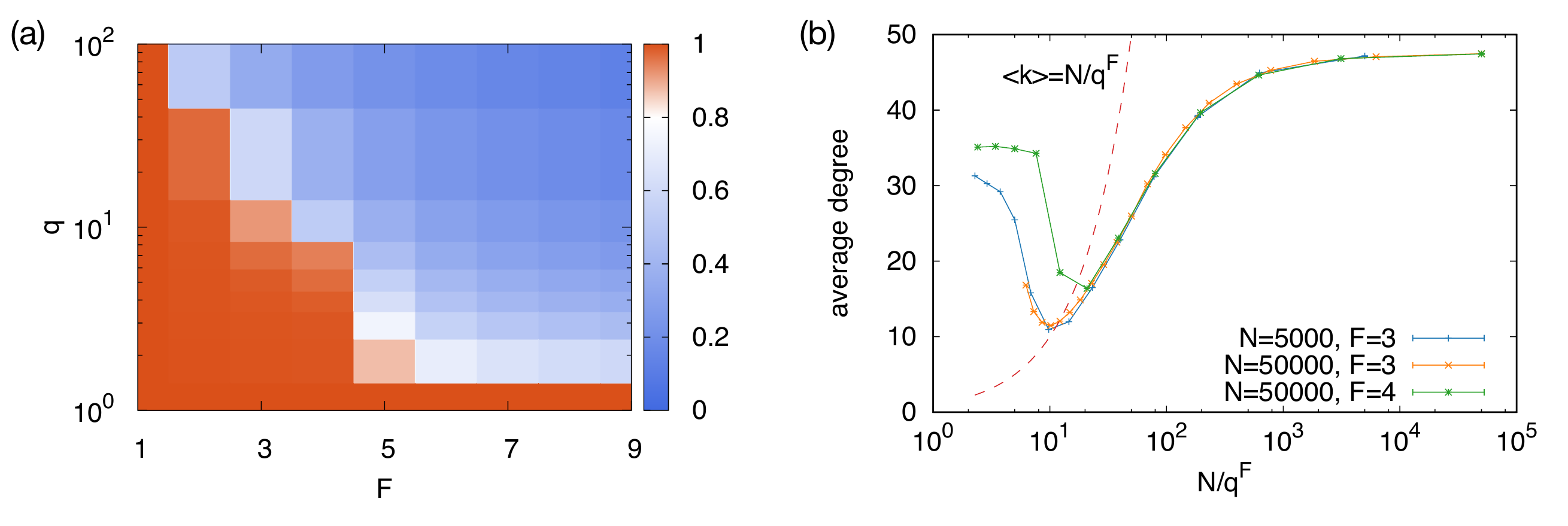}
  \caption{
  (a)~Feature overlap as a function of $F$ and $q$.
  In the segregated phase, feature overlap shows a value close to $1$ while it is significantly less than $1$ in the overlapping phase.
  When $F=1$ or $q=1$, the feature overlap is $1$ by model definition.
  (b)~Scaling plot of the average degree as a function of the system size $N$, the number of features $F$, and the number of different values $q$ each feature can take. For $F$ around $F_c$, we get good scaling with the variable $N/q^F$. The dashed line corresponds to $\langle k\rangle =N/q^F$, which predicts the transition points. Note that some deviations around the minimum may be due to the discreteness of $F$.
  }
  \label{fig:phase_diagram_scaling}
  \end{center}
\end{figure}

\section*{Results}\label{sec:results}

First, we show the dependence of the average degree on $F$ and $q$ in Fig.~\ref{fig:k_F}(a). For $q>1$ we find two regimes: The average degree decreases as $F$ increases up to a certain value, $F_c$, then it increases for $F>F_c$. Such non-monotonic dependence is also observed for the clustering coefficient in Fig.~\ref{fig:k_F}(b). The clustering coefficient shows increasing and decreasing behaviors for $F<F_c$ and $F>F_c$, respectively, hence it has a peak at around $F_c$, implying that the network structure qualitatively changes at around $F_c$.

The transition point $F_c$ depends on the WSN parameters but, more importantly, also on $q$; as it is clear from Fig.~\ref{fig:k_F} the function $F_c(q)$ is decreasing. Consequently, for a given $F$ a similar transition takes place as a function of $q$ thus one can define $q_c(F)$. 

To investigate what is happening at the transition point, we measure how many of the features are shared between the neighboring nodes. We define the feature overlap between nodes $i$ and $j$ as the fraction of shared features:
\begin{equation}
o_F(i,j) \equiv \frac{1}{F}\sum_{f=1}^{F} \delta(\sigma_i^f,\sigma_j^f),
\end{equation}
where $\delta(x,y)$ is the Kronecker delta. Figure~\ref{fig:overlap_F}(a) shows the feature overlap averaged over the connected node pairs, $\langle o_F \rangle$, as a function of $F$. $\langle o_F \rangle$ stays close to $1$ in the low $F$ phase indicating that most of the links are created between the nodes having completely identical features. Hereafter, we denote the nodes having $o_F=1$ as ``matching nodes''. On the other hand, as $F$ becomes greater than $F_c$, the feature overlap quickly drops from unity and converges to a value slightly greater than $1/q$. Due to the above results we denote the two regimes of $F<F_c$ and $F>F_c$ as ``segregated'' and ``overlapping'' phases, respectively, since nodes are segregated into loosely-connected homogeneous communities in the former phase, while the majority of the communities overlap with each other in the latter phase. The corresponding phase diagram is shown in Fig.~\ref{fig:phase_diagram_scaling}(a).

Let us shed light onto the origin of the transition. Since the feature values are drawn from a uniform discrete distribution from $\{1,\cdots,q\}$, the number of nodes sharing a specific set of traits is on average $\tilde{N} \approx N/q^F$. The transition occurs when $\tilde{N}$ becomes less than the typical degree for the original WSN model. For instance, the transition point $F_c=4$ for $q=5$ leads to $\tilde{N}=5\times 10^4/5^4 = 80$, which is of the order of the average degree $47$ for the original WSN model with the same parameters. Therefore, we expect that $N/q^F$ is a scaling variable for the average density, see Fig.~\ref{fig:phase_diagram_scaling}(b).

By the model definition, each node has a strong tendency to connect to matching nodes as long as a matching node can be found, i.e., in the segregated phase. In the overlapping phase, on the other hand, the chance to find matching nodes becomes small thus the feature overlap cannot be one anymore. The nodes compromise with partially matching nodes as few matching nodes exist in the whole population.

When $F<F_c$, the average degree decreases as a function of $F$ as shown in Fig.~\ref{fig:k_F}(a). This is because the number of matching nodes $\tilde{N}$, which is equal to the number of potential neighbors, decreases with $F$. The opposite behavior is found in the other phase, where the nodes are connected to others that share at least one feature. The number of nodes sharing at least one feature is calculated as $N[1-(1-1/q)^F]$, which is an increasing function of $F$. Since there is no chance for building communities of matching nodes for $F>F_c$ the nodes have more potential neighbors with increasing $F$ thus they will have higher degrees. The spike found in the clustering coefficient [Fig.~\ref{fig:k_F}(b)] can also be explained similarly. When $F$ gets closer to $F_c$ in the segregated phase, the number of potential neighbors decreases and eventually becomes the same order as the degree. Since nodes are partitioned into groups of matching nodes and its size is comparable to the average degree, the nodes within a group are mostly connected to each other, while the number of links bridging the groups is small. Thus, the network is mostly composed of $q^F$ clique-like components, yielding a high clustering coefficient as well as a high assortativity (not shown).

This transition behavior cannot be simply explained by the naive assumption that the probability of making a link is proportional to $o_F$. Let us consider a null model in which links are randomly created between nodes $i$ and $j$ with a probability proportional to $o_F(i,j)$. The expected number of links between the node pairs having $o_F=n/F$ is given by the binomial form as
\begin{equation}
L\left(o_F=\frac{n}{F}\right) = \frac{cn}{F}\frac{N(N-1)}{2} \binom{F}{n}\left(\frac{1}{q}\right)^n\left(1-\frac{1}{q}\right)^{F-n},
\end{equation}
where $c$ is a coefficient for the probability of creating links. Therefore, the average feature overlap in the network is calculated as
\begin{eqnarray}
\langle o_F \rangle = \frac{\sum_{n=1}^{F} nL(o_F=n/F)/F }{ \sum_{n=1}^{F} L(o_F=n/F) } 
   =  \frac{1}{q} + \frac{1}{F}\left(1-\frac{1}{q} \right).
\end{eqnarray}
This null model predicts a smooth monotonic decrease without a transition with increasing $F$ [see Fig.~\ref{fig:overlap_F}(a)], which is significantly different from the simulation results and with which it agrees only asymptotically.

Next we test a model with GA and LD but without LA by setting $p_{\Delta}=0$. The results in Fig.~\ref{fig:overlap_F}(a) do not show a transition and reproduce a curve similar to the null model. Thus, a key ingredient for the transition is the presence of LA rule. In other words, LA promotes a much stronger preference in link formation between matching pairs than predicted by the null model. Indeed, the link reinforcement in LA plays here a key role as initially minor differences between the probabilities of choosing neighbors with perfect matching and partial overlaps get strongly amplified due to the positive feedback produced by the weight-dependence at the selection of the neighbors. As a consequence, the links created by triadic closures are almost always between matching nodes in the segregated phase. This argument has been numerically confirmed by disabling the link reinforcement by setting $w_r=0$ while keeping a positive $p_{\Delta}$, as shown in Fig.~\ref{fig:overlap_F}(a).

While the above amplification effect exists in the overlapping phase too, it cannot fully develop as the low number of available matching nodes results in a partial overlap. For instance, even when nodes $A$ and $B$ are both connected to $C$ and differ in their traits from those of $C$'s in one feature only, when $A$ and $B$ get connected by cyclic closure their traits may differ already in two features. Thus, the feature overlap between neighbors significantly decreases when nodes cannot connect to matching nodes.

\begin{figure}
  \begin{center}
  \includegraphics[width=.5\textwidth]{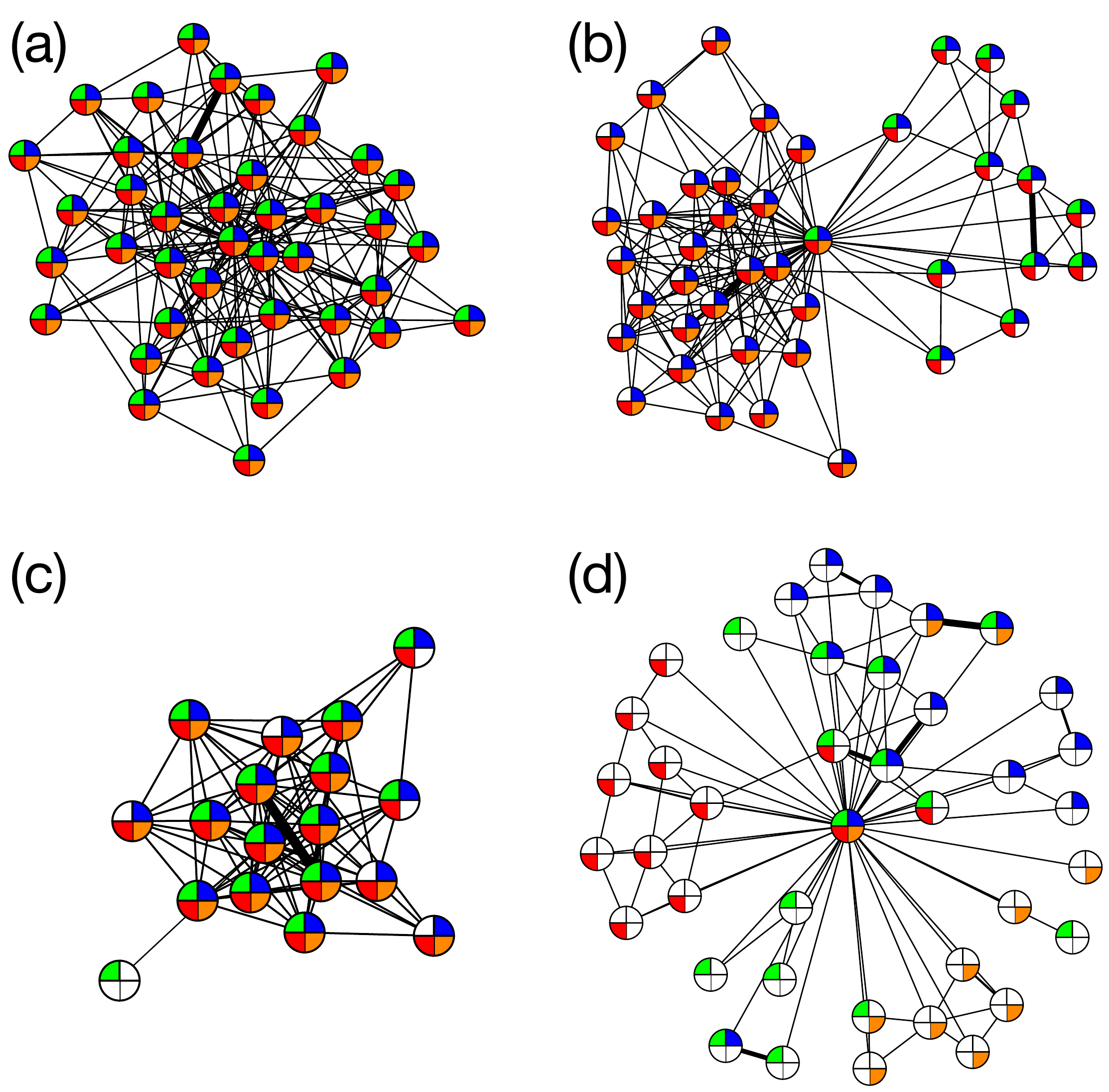}
  \caption{
  Typical egocentric networks in the case with $F=4$ for (a,~b)~$q=4$ (segregated phase), (c)~$q=7$ (near the transition point), and (d)~$q=20$ (overlapping phase). The central node denotes the ego. When a feature of an acquaintance is identical to that of the ego it is colored otherwise left white. In the segregated phase, a node is typically connected to its matching nodes~(a) while a small fraction of nodes have links to a few other types of nodes, bridging communities~(b).
  (c)~A node has a smaller degree, and some of its neighbors are not matching nodes.
  (d)~Feature overlap between neighboring nodes becomes even less and the ego belongs to more diverse communities.
  }
  \label{fig:ego_net}
  \end{center}
\end{figure}

The fact that $\langle o_F \rangle \approx 1$ in the segregated phase indicates that the nodes in the network form $q^F$ communities, each of which mostly consisting of matching nodes. Whether this segregation sets in or not depends on the value of $F$: At $F_c$ the network structure changes from large homogeneous, segregated clusters to smaller more heterogeneous groups. This transition can be directly observed by applying the community detection to the generated networks. We use the multi-level Infomap algorithm~\cite{rosvall2011multilevel} to detect a hierarchical structure of the communities. Since we are interested in the global segregation, we pick up the communities on the top of the hierarchy, and their overlap is measured by the number of communities per node as shown in Fig.~\ref{fig:overlap_F}(b) as a function of $F$. We clearly see two distinct regimes. When $F<F_c$, we have homogeneous communities consisting mainly of matching nodes thus the number of communities per node is close to one, indicating that there is little overlap of the membership between communities. For $F>F_c$ the feature overlap between linked nodes is suppressed. As a consequence, the membership overlap between communities increases, which works against segregation. This is why we named the two phases as segregated and overlapping phases.

These differences are illustrated in Fig.~\ref{fig:ego_net}(a) where the egocentric network of a typical node is shown, and clearly all acquaintances have identical traits, i.e., matching nodes. There are, however, some ($\sim 5\%$) nodes which connect these large homogeneous communities. %of the size of $~N/q^F$. 
A typical example of these connecting nodes is shown in Fig. ~\ref{fig:ego_net}(b). Near the transition point, as shown in Fig.~\ref{fig:ego_net}(c), a node typically has a smaller degree and shares less feature with its neighbors than those in the segregated phase. Figure~\ref{fig:ego_net}(d) shows a sample from the overlapping regime. The ego is a part of more distinct communities that have small feature overlap because of a large $q$, as shown in Fig.~\ref{fig:overlap_F}(a).
%ve decreasing feature overlap as $q$ increases as shown in 
As nodes are part of more and more communities the number of communities per node increases as shown in Fig.~\ref{fig:overlap_F}(b).

\begin{figure}
  \begin{center}
  \includegraphics[width=.48\textwidth]{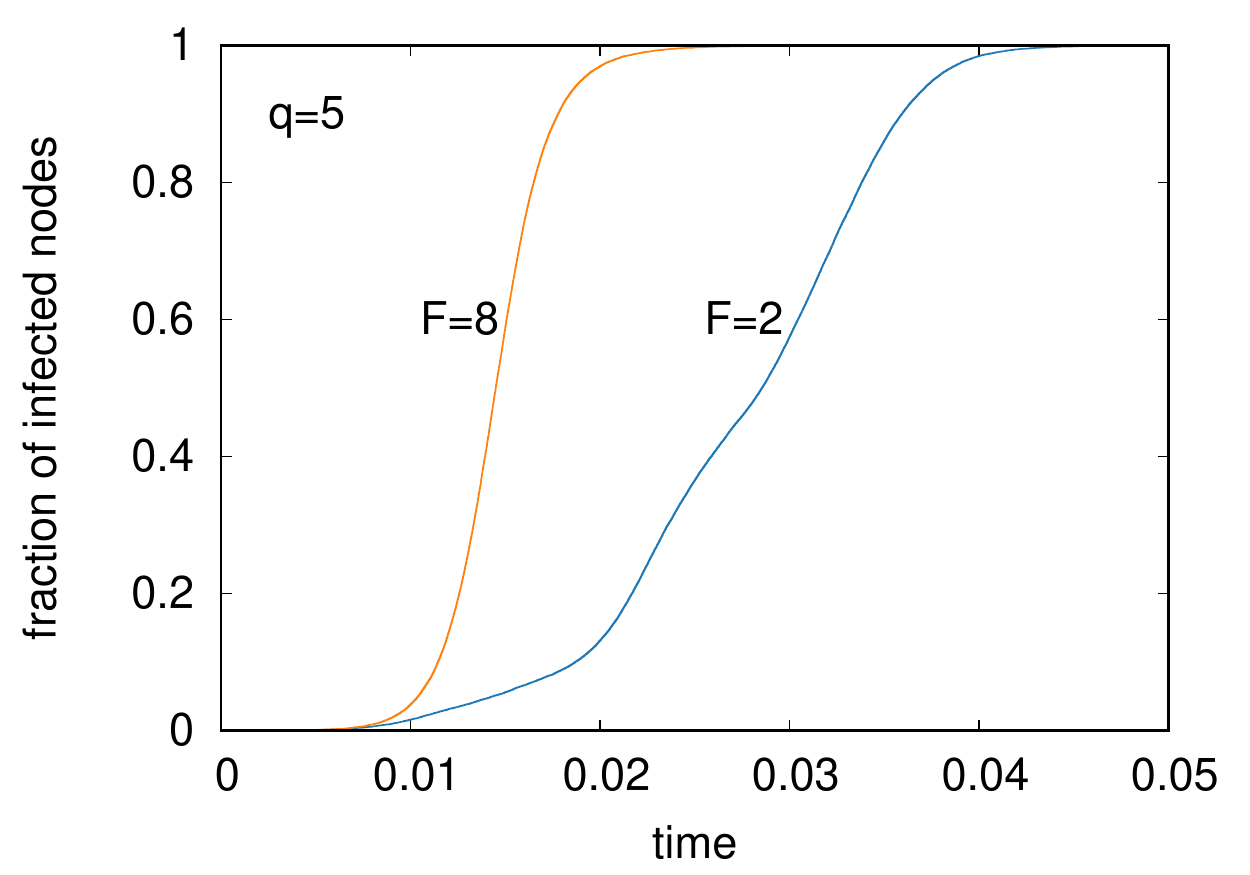}
  \caption{
  Information spreading on the networks obtained for $F=2$ (segregated phase) and $8$ (overlapping phase) when $q=5$.
  Simulations of the SI model are conducted where the spreading probability is assumed to be proportional to the link weights.
  Note that the unit of the horizontal axis is arbitrarily scaled without loss of generality.
  The infection starts from a randomly chosen node and continues until all the nodes are infected.
  Even though these two networks have similar degree and average link weight, the spreading speed in the overlapping phase is approximately two times faster than for the segregated phase.
  These are the results of single Monte Carlo runs.
  }
  \label{fig:spreading}
  \end{center}
\end{figure}

Another way to investigate the segregation is to see information spreading on the network. We simulate the susceptible-infected (SI) dynamics~\cite{PastorSatorras2015Epidemic} on the networks assuming that the transmission probability is simply proportional to link weights, irrespective of the features of nodes. The information starts to spread from a randomly chosen node and the fraction of infected nodes are calculated until all the nodes become infected. The SI model is able to detect the influence of the topology and the link weights onto the spreading speed~\cite{onnela2007structure, Karsai2011small}. Figure~\ref{fig:spreading} shows typical dynamics of spreading on the networks for $F=2$ (segregated phase) and $F=8$ (overlapping phase) when $q=5$. These two networks have been chosen because they have similar average degrees ($46.5$ and $47.4$, respectively) and average link weights ($21.8$ and $21.3$, respectively). Although these two networks have comparable local network statistics, the spreading for the segregated phase is much slower than that for the overlapping phase. In the segregated phase, the spreading is trapped in a community of matching nodes: It takes a longer time to convey information to a node having different features than for the overlapping phase, in which spreading between different groups occurs more easily.

\section*{Summary and Discussion}\label{sec:discussion}

In this paper we have generalized the weighted social network model to incorporate homophily and to study its effect on the emerging network structure. We introduced $F$ features for each node and every feature could have $q$ different values. The GA and LA steps were made then dependent on the overlap between the nodes' features. The model shows two phases depending on the parameters $F$ and $q$. For a given value of $q$ and $F<F_c$ the nodes in the whole network are segregated into non-overlapping communities, consisting mostly of matching nodes. On the other hand, when $F>F_c$, such segregation is not observed. Nodes partially share features with their neighbors yielding significantly larger amount of overlapping communities. Thus, if the selection mechanism of acquaintances is based on very few features ($F$ is small) and/or if traits a feature can take are limited only to few options ($q$ is small), then the society will be segregated. 

What is the relevance of this finding, considering that in reality every person has a large number of features? We note that the features are not equally important in tie formation. There are cultures and situations, in which a few features get extreme importance. This is the case, in times of sharpening political situations, turmoils or wars. Another relevant example can be social groups in the online social networks, where people get into contact based on very few features. This (especially if amplified by the mentioned political conditions) leads then to the well-known effect of ``echo chambers''~\cite{Barbera2015tweeting} and our results indicate that perceiving the complexity of the world or allowing for more subtle opinions can break up these bubbles.

The overlapping phase is thus a more realistic model of a normal society where one has acquaintances of different gender, age, location, political view, sport preference, etc. Each of these traits has its own communities in which there is no perfect feature overlap. For example, why should we do sports with people of the same political preference?

We demonstrated that the feature similarity between nodes can be much more amplified if the local attachment mechanism is taken into account. In other words, the homophily principle and the local attachment mechanism including reinforcement have a joint effect in increasing segregation in social networks. Our study implies strategies to avoid segregation and splitting of the society by maintaining cultural diversity.

The relationship between homophily and segregation has been recognized long ago. In the Schelling-type models~\cite{schelling1969models, henry2011emergence} people change their position in a fixed geometry (usually a lattice) to better satisfy their preferences. In another approach to social segregation, opinion dynamics is used so that similar people can influence each other~\cite{axelrod1997dissemination, centola2007homophily, vazquez2007time, deffuant2000mixing, del2017modeling}. In these models similarity is amplified by the changes in the opinions (or features). The model proposed in this paper has a different mechanism. Our aim was to investigate how homophily influences the social network if such elementary tie formation mechanisms like focal and cyclic closure (mimicked by our model as GA and LA, respectively) are taken into account. Our conclusion is that social segregation may be intensified by LA and link reinforcement even without changing the features in interactions. For more realistic modeling, we can of course think of a combined model incorporating both the feature updates and the triadic closure with reinforcement. Although we leave it for future studies, we expect that the segregation is observed in a wider range of parameter space as the feature updates promote segregation in general.

There are many further ways to extend our model and make it more realistic. A possible extension is to allow different $q$ values for each feature. This is a much more realistic setting and will be necessary, when comparing with a data. Also, individual differences in the number of features could be considered. 

\bibliography{complex-network}

%\noindent LaTeX formats citations and references automatically using the bibliography records in your .bib file, which you can edit via the project menu. Use the cite command for an inline citation, e.g.  \cite{Hao:gidmaps:2014}.

%For data citations of datasets uploaded to e.g. \emph{figshare}, please use the \verb|howpublished| option in the bib entry to specify the platform and the link, as in the \verb|Hao:gidmaps:2014| example in the sample bibliography file.

\section*{Acknowledgements}

Y.M. acknowledges support from MEXT as “Exploratory Challenges on Post-K computer (Studies of multi-level spatiotemporal simulation of socioeconomic phenomena)” and from Japan Society for the Promotion of Science (JSPS) (JSPS KAKENHI; grant no. 18H03621).
H.-H.J. acknowledges financial support by Basic Science Research Program through the National Research Foundation of Korea (NRF) grant funded by the Ministry of Education (NRF-2018R1D1A1A09081919). K.K. acknowledges the Rutherford Foundation Visiting Fellowship at The Alan Turing Institute, UK. The systematic simulations in this study were assisted by OACIS~\cite{murase2017open}. Y.M., H.-H.J., J.T., and J.K. are thankful for the hospitality of Aalto University.

\section*{Author Contributions}

Y.M., H.-H.J., J.T., J.K. and K.K. conceived and designed the experiments.
Y.M. and J.T. conducted the experiments and analyzed the results.
Y.M., H.-H.J., J.T., J.K. and K.K. authors reviewed the manuscript.

\section*{Data Availability}
All data generated or analyzed during this study are included in this published article.

\section*{Competing Interests}
The authors declare no competing interests.

\end{document}